\documentclass{PoS}

\PoS{PoS(LAT2005)045}

\title{Scaling test of the P4-improved staggered fermion action. }

\ShortTitle{Scaling test of the P4-improved staggered fermion action. }

\author{\speaker{Michael Cheng} for the RBC-Bielefeld Collaboration \\
        Columbia University\\
        E-mail: \email{michaelc@phys.columbia.edu}}

\abstract{ We present a scaling study of the QCD spectrum using a
smeared P4 staggered fermion formulation, in which three, five, and 
seven-link staples are added to reduce the effects of
flavor symmetry breaking.  These studies are performed on quenched 
lattices generated using the one-loop improved Symanzik gauge 
action, with $\beta=\frac{10}{g^2}=7.40,7.75,8.00$.  The corresponding 
lattice spacings are a = .31 fm, .21 fm, and .14 fm.  Particularly,
we study the $\mathit{O}(a^2)$ flavor symmetry violations in the
pion spectrum as well as the scaling dependence of $m_{\rho}$, measured
in physical units, on lattice spacing.
These results are compared against previous results for the Asqtad action.}

\FullConference{XXIIIrd International Symposium on Lattice Field Theory\\
                 25-30 July 2005\\
                 Trinity College, Dublin, Ireland}
\begin{document}

\section{Introduction}
Lattice QCD is an important method for computing interesting quantities
in the Standard Model as well as in finite temperature QCD.  
However, strictly speaking, the different lattice formulations of QCD 
reproduce the continuum theory only in the limit that the lattice spacing 
goes to zero ($a \rightarrow 0$).  Thus, it is important to choose a lattice 
action where the finite lattice spacing errors are as small as possible.
One of the more popular improved fermion actions is the Asqtad action
\cite{Orginos:1999cr}.  It has been shown from quenched measurements of 
various quantities that the Asqtad action has much smaller
scaling errors than many other unimproved actions\cite{Bernard:1999xx}.

Another interesting improved action is the P4 action\cite{Heller:1999xz}.  
In these proceedings we study the scaling properties of the P4 action, 
particularly as compared to Asqtad.  To accomplish this, we make quenched 
measurements of the Goldstone and non-Goldstone pion masses.  The mass 
splitting between these two particles is one manifestation of 
$\mathit{O}(a^2)$ effects in the staggered formulation\cite{Lee:1999zx}.  
We also look at how $m_{\rho}$ scales with lattice spacing, and the effect 
of tadpole improvement\cite{Lepage:1992xa} on all of these quantities.
\vspace{-2mm}

%%%%%%%%%%%%%%%%%%%%%%%%%%%%%%%%%%%%%%%%%%%%%%%%%%%%%%%%%%%%%%%%%%%%%%%%%%%%%

\section{P4 action}
Like the Asqtad action, the P4 action is a variant of the Kogut-Susskind
formulation of lattice fermions.  To improve the naive staggered action, 
we allow the inclusion of higher order derivatives.  There are two possible 
three-link terms allowed by the hypercubic symmetry of the staggered 
formulation.  These are the familiar Naik term\cite{Naik:1986bn}, and the 
so-called ``knight's move'' term:
\vspace{-1mm}
\begin{eqnarray}
  S_F & = &       m\sum_{x}\bar{\chi}(x)\chi(x) + 
  \sum_{x}\bar{\chi}(x)\sum_{\mu}\eta_{\mu}(x)\nonumber\\
       & &     \Big\{ c_{1,0} \Big[U_{\mu}(x)\chi(x+\mu) - U_{\mu}^{\dagger}
	      (x-\mu) \chi(x-\mu)\Big] + {} \\
       & &  + c_{3,0}\Big[U_{\mu}^{(3,0)}(x)\chi(x+3\mu)-U_{\mu}^{(3,0)\dagger}
            (x-3\mu)\chi(x-3\mu)\Big] + {} \nonumber \\
      & & + c_{1,2}\sum_{\nu\neq\mu}\Big[U_{\mu,\nu}^{(1,2)}(x)
               \chi(x+\mu+2\nu)- U_{\mu,\nu}^{(1,2)\dagger}(x-\mu-2\nu) 
               \chi(x-\mu-2\nu)+ {}\nonumber\\
      & & \qquad \qquad + U_{\mu,\nu}^{(1,-2)}(x)\chi(x+\mu-2\nu)
	  -U_{\mu,\nu}^{(1,-2)\dagger}(x-\mu+2\nu)\chi
          (x-\mu-2\nu)\Big] \Big\} \nonumber
\end{eqnarray}
\vspace{-1mm}
where $\chi(x)$ is the one-component fermion field, $\eta_{\mu}(x)$ is
the staggered phase, and the various parallel transporters are given by
\vspace{-1mm}
\begin{eqnarray}
U_{\mu}^{(3,0)}(x) & = & U_{\mu}(x)U_{\mu}(x+\mu)U_{\mu}(x+2\mu) \\
U_{\mu,\nu}^{(1,2)}(x) & = & \frac{1}{2}\big[ U_{\mu}(x)U_{\nu}(x+\mu)
U_{\nu}(x+\mu+\nu)+U_{\nu}(x)U_{\nu}(x+\nu)U_{\mu}(x+2\nu)\big] \nonumber \\
U_{\mu,\nu}^{(1,-2)}(x) & = & \frac{1}{2}\big[ U_{\mu}(x)U_{\nu}^{\dagger}
(x+\mu-\nu)U_{\nu}^{\dagger}(x+\mu-2\nu)+U_{\nu}^{\dagger}(x-\nu)
U_{\nu}^{\dagger}(x-2\nu)U_{\mu}(x-2\nu)\big] \nonumber
\end{eqnarray}
\vspace{-1mm}
If we examine the naive quark propagator ($c_{1,0} = 1, c_{3,0} = 0, 
c_{1,2} = 0$), we see that rotational symmetry is violated at $\mathit{O}
(p^4)$.  However, with an improved staggered action, we can systematically
eliminate these violations if the coefficients satisfy the constraint
$c_{1,0}+27c_{3,0}-18c_{1,2} = 0$.
Coupled with an overall continuum normalization, this produces a one-parameter
family of actions that remove the leading-order rotational symmetry violation
in the quark propagator.  The improvement coefficients satisfy
$c_{1,2}-c_{3,0} = \frac{1}{24}$.

If we set $c_{1,2}=0, c_{3,0}=-\frac{1}{24}$, we recover the Naik action.
The choice $c_{1,2}=\frac{1}{24}, c_{3,0}=0$ yields the P4 action.
Not only does the Naik action improve the rotational symmetry, but it
also removes all $\mathit{O}(a^2)$ errors in the staggered propagator
\cite{Naik:1986bn}.  However, if one examines the free-field dispersion 
relation, one sees that it is the P4 action that more effectively removes 
higher-order corrections, thus providing a much better approximation to the 
free-field, continuum dispersion relation\cite{Heller:1999xz}.
\begin{table}
{\large
\begin{center}
  \begin{tabular}{cccccccc}
    Action & $c_1$ & $c_{Naik}$ & $c_{knight}$ & $c_3$ & $c_5$ & $c_7$ 
    & $c_{Lepage}$\\
    \hline
    \emph{Asqtad} & $\frac{1}{8}+\frac{3}{8}+\frac{1}{8}$ & 
    $-\frac{1}{24u_0^2}$ & $0$ & $\frac{1}{16u_0^2}$ & $\frac{1}{64u_0^4}$ 
	 & $\frac{1}{384u_0^6}$ & $-\frac{1}{16u_0^4}$\\
    \emph{P4fat7} & $\frac{1}{8}-\frac{1}{4}$ & $0$ & $\frac{1}{24}$ &
    $\frac{1}{16}$ & $\frac{1}{64}$ & $\frac{1}{384}$ & $0$\\
    \emph{P4fat7tad}& $\frac{1}{8}-\frac{1}{4}$ & $0$ & $\frac{1}{24u_0^2}$ &
    $\frac{1}{16u_0^2}$ & $\frac{1}{64u_0^4}$ & $\frac{1}{384u_0^6}$ & $0$\\
  \end{tabular}
\end{center}
}
\vspace{-5mm}
\caption{\label{actioncoeff}Fermion action coefficients.  $c_{Lepage}$
is introduced in the Asqtad action
to remove the $\mathit{O}(a^2)$ errors introduced by fat link smearing
\cite{Lepage:1998vj}.  We
choose not to include it in either variant of the P4 action.}
\end{table}
\vspace{-2mm}

%%%%%%%%%%%%%%%%%%%%%%%%%%%%%%%%%%%%%%%%%%%%%%%%%%%%%%%%%%%%%%%%%%%%%%%%%%%%%%

\section{Smearing and Tadpole Improvement}
The Asqtad action incorporates not only the Naik term, but also makes
use of fat-link smearing\cite{Lepage:1997id} with tadpole-improved 
coefficients to reduce 
the effects of flavor-changing interactions.  In order to take full
advantage of any possible flavor-symmetry improvement, we incorporate
fat-link smearing into our versions of the P4 action.
Specifically, we replace the one-link part of the action with the
combination
\vspace{-3mm}
\begin{center}
\includegraphics[width=0.6\textwidth]{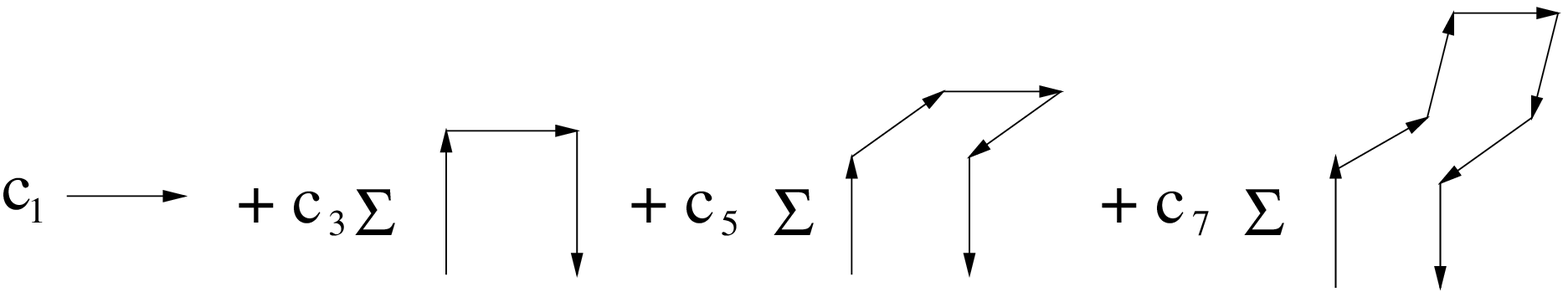}
\end{center}
\vspace{-2mm}
The smearing coefficients are tuned to remove flavor-changing couplings
to lowest order in perturbation theory.  Furthermore, in order to study
the effect of tadpole improvement, we  
measure the hadron spectrum for two different variants of the P4 action.
What we call \emph{P4fat7} utilizes fat-link smearing without tadpole 
improvement, while \emph{P4fat7tad} uses tadpole-improved coefficients.
Consult Table \ref{actioncoeff} for a full list of parameters.
\begin{table}
{\small
\begin{center}
 \begin{tabular}{|cccccc|}
   \hline
   $\beta = 10/g^2$ & $a[fm.]$ & $r_1/a$ & plaquette & $plaq^{\frac{1}{4}}$ & input $u_0$ \\
   \hline
   7.40 & .31 & 1.44(1) &.554799(6) & .863045(2) & .8629 \\
   7.75 & .21 & 2.08(5) &.599179(7) & .879811(2) & .8800  \\
   8.00 & .14 & 2.65(1) &.621287(4) & .887816(1) & .8879 \\
   \hline
   \end{tabular}
  \end{center}}
\vspace{-3mm}
\caption{\label{simpar} Lattice spacing $a$, string tension scale $r_1$, 
and plaquette for the quenched evolutions}
\vspace{-3mm}
\end{table}
\vspace{-2mm}

%%%%%%%%%%%%%%%%%%%%%%%%%%%%%%%%%%%%%%%%%%%%%%%%%%%%%%%%%%%%%%%%%%%%%%%%%%%%%

\section{Simulation details}
Spectrum measurements were done for the Asqtad, P4fat7, and P4fat7tad actions
on three different sets of quenched lattices, chosen to match the lattices
used by MILC in their Asqtad scaling study\cite{Bernard:1999xx}.  
Lattices were generated using
the one-loop tadpole-improved Symanzik gauge action with a gauge heat bath
evolution.  We chose $\beta = \frac{10}{g^2} =$ 7.4, 7.75, and 8.00 with 
volumes of $16^3\times32$ for the first two ensembles and $24^3\times32$ for the latter ensemble.  

For the $16^3\times32$ lattices, spectrum measurements were made on
100 configurations, separated by 1000 heat bath sweeps.  For the $24^3\times32$
lattices, 100 measurements were made on lattices separated by 500 heat
bath sweeps.  The hadron spectrum was measured for each fermion action
for a set of at least three masses using a $2\mathbb{Z}$ wall source.
All computation was done on QCDOC computers.  Table \ref{simpar} shows the 
values for the plaquette, the string tension scale $r_1$\cite{Bernard:1999xx}, 
and the tadpole-improvement coefficient $u_0$.
\begin{table}
  {\small
  \begin{center}
 \begin{tabular}{|c|c|cccc|cccc|}
   \hline
   & & \multicolumn{4}{|c|}{$\beta=7.40$} & \multicolumn{4}{|c|}{$\beta=7.75$}\\
   \cline{3-10}
   Action  & $m_{q}a$ & $m_{\pi}a$ & $m_{\pi_2}a$ & $m_{\rho}a$ & $m_{\rho_2}a$
   & $m_{\pi}a$ & $m_{\pi_2}a$ & $m_{\rho}a$ & $m_{\rho_2}a$\\
   \hline
   P4fat7  & .01 & .4210(3) & .672(9) & 1.06(2) & 1.09(5)
   & .3221(5) & .397(3) & .777(12) & .803(19)\\
   & .02 & .5842(3) & .797(4) & 1.13(1) & 1.17(1)
   & .4500(6) & .505(2) & .819(7) & .835(7)\\
   & .025 &-&-&-&-& .5015(6) & .5512(14) & .843(5) & .854(5)\\
   & .03 & .7050(3) & .895(3) & 1.199(5) & 1.231(7)
   & .5482(7) & .5954(12) & .871(4) & .883(6)\\
   \hline
   P4fat7tad  & .02 & .4229(3) & .634(7) & 1.06(2) & 1.06(4)
   &.3467(5) & .409(2) & .766(20) & .801(10)\\
   & .025 &-&-&-&-& .3861(5) & .443(2) & .786(12) & .813(11)\\
   & .03 & .5143(3) & .707(4) & 1.10(1)  & 1.11(2)
   & .4217(6) & .477(2) & .810(15) & .820(8)\\
   & .04 & .5905(4) & .766(3) & 1.132(6) & 1.148(7)
   & .5015(6) & .5512(14) & .843(5) & .854(5)\\
   \hline
   Asqtad  & .02 & .3782(3) & .668(9) & 1.14(2) & 1.10(2)
   & .3307(5) & .422(2) & .793(14) & .819(18)\\
   & .025 &-&-&-&-& .3686(5) & .4511(18) & .804(10) & .841(20)\\
   & .03 & .4606(3) & .719(5) & 1.16(1) & 1.15(3)
   & .4027(5) & .481(3) & .813(12) & .838(12)\\
   & .04 & .5292(4) & .767(4) & 1.18(1) & 1.18(2)
   & .4631(7) & .5310(16) & .841(8) & .860(8)\\
   \hline
   & &\multicolumn{4}{|c|}{$\beta=8.00$}\\
   \cline{3-6}
   Action  & $m_{q}a$ & $m_{\pi}a$ & $m_{\pi_2}a$ & $m_{\rho}a$ & $m_{\rho_2}a$\\
   \cline{1-6}
   P4fat7  & .01 & .2736(5) & .3026(9) & .627(6) & .625(7)\\
   & .02 & .3821(6) & .4025(6) & .666(3) & .670(3)\\
   & .03 & .4662(5) & .4827(5) & .707(2) & .710(2)\\
   \cline{1-6}
   P4fat7tad  & .01 & .2170(5) & .2519(13) & .605(7) & .605(8)\\
   & .02 & .3025(5) & .3270(8) & .635(5) & .634(5)\\
   & .03 & .3679(4) & .3876(6) &  .659(3) & .661(3)\\
   \cline{1-6}
   Asqtad  & .01 & .2107(4) & .2558(15) & .619(8) & .607(7)\\
   & .02 & .2941(4) & .3272(8) & .648(5) & .637(5)\\
   & .03 & .3579(6) & .3856(8) & .669(3) & .667(5)\\
   \cline{1-6}
 \end{tabular}
  \end{center}}
  \caption{\label{masses} Hadron masses for $\beta=7.40$, $\beta=7.75$, and
  $\beta=8.00$}
  \vspace{-3mm}
\end{table}
\vspace{-3mm}

%%%%%%%%%%%%%%%%%%%%%%%%%%%%%%%%%%%%%%%%%%%%%%%%%%%%%%%%%%%%%%%%%%%%%%%%%%%%

\section{Results}
\vspace{-1mm}
Table \ref{masses} show the result of our spectrum measurements.  For each
value of $\beta$ = $7.40, 7.75,$ and $8.00$, we measured the mass for
the pseudoscalar meson ($m_{\pi}$), the scalar meson ($m_{\pi_2}$), the
vector meson ($m_{\rho}$), and the pseudovector meson ($m_{\rho_2}$).  As
expected, we can see the $\mathit{O}(a^2)$ splitting between the pseudoscalar
and scalar pion at all lattice scales.
\begin{figure}[hbt]
  \vspace{-5mm}
  \includegraphics[angle=-90, width=.53\textwidth]{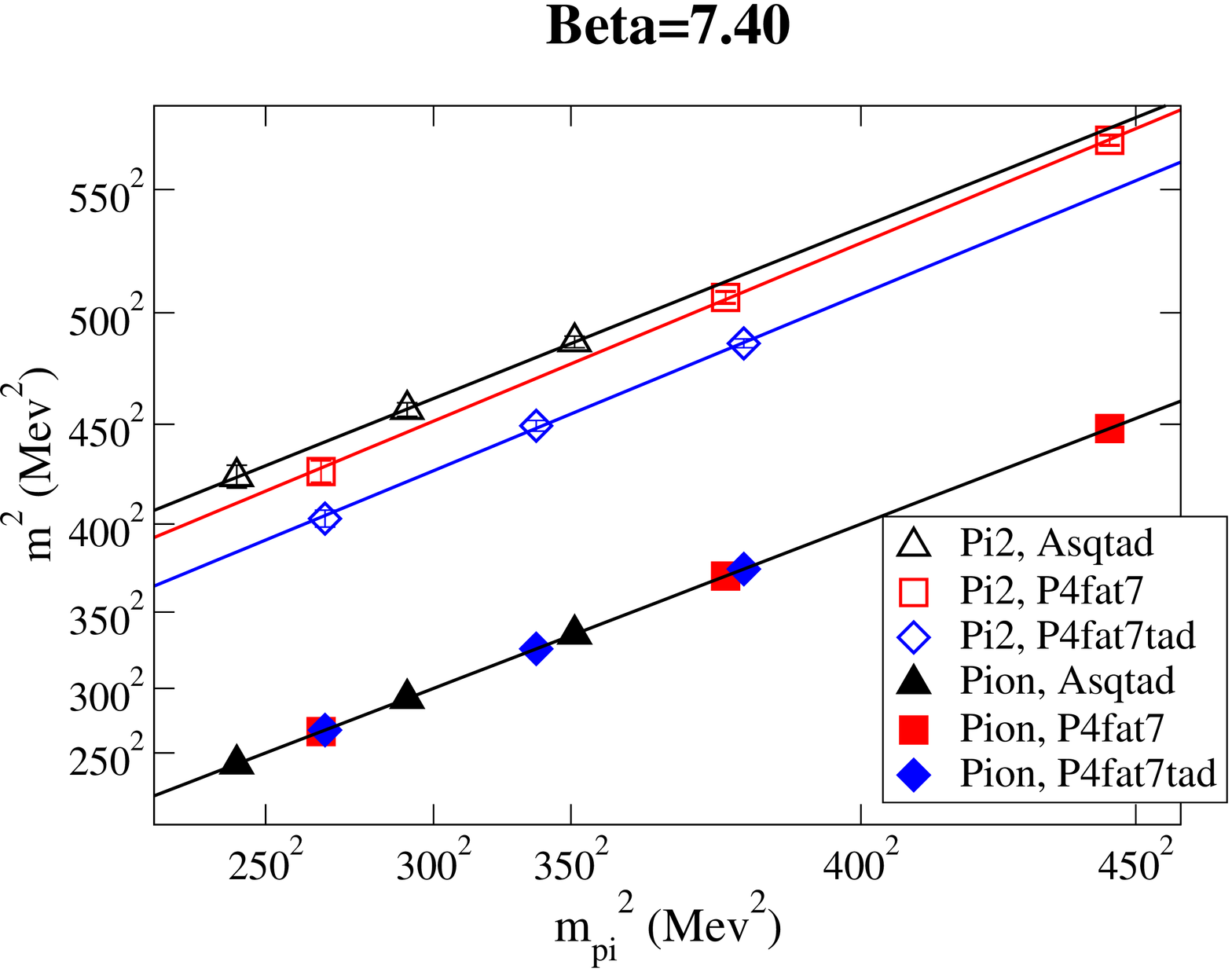}
  \includegraphics[angle=-90, width=.53\textwidth]{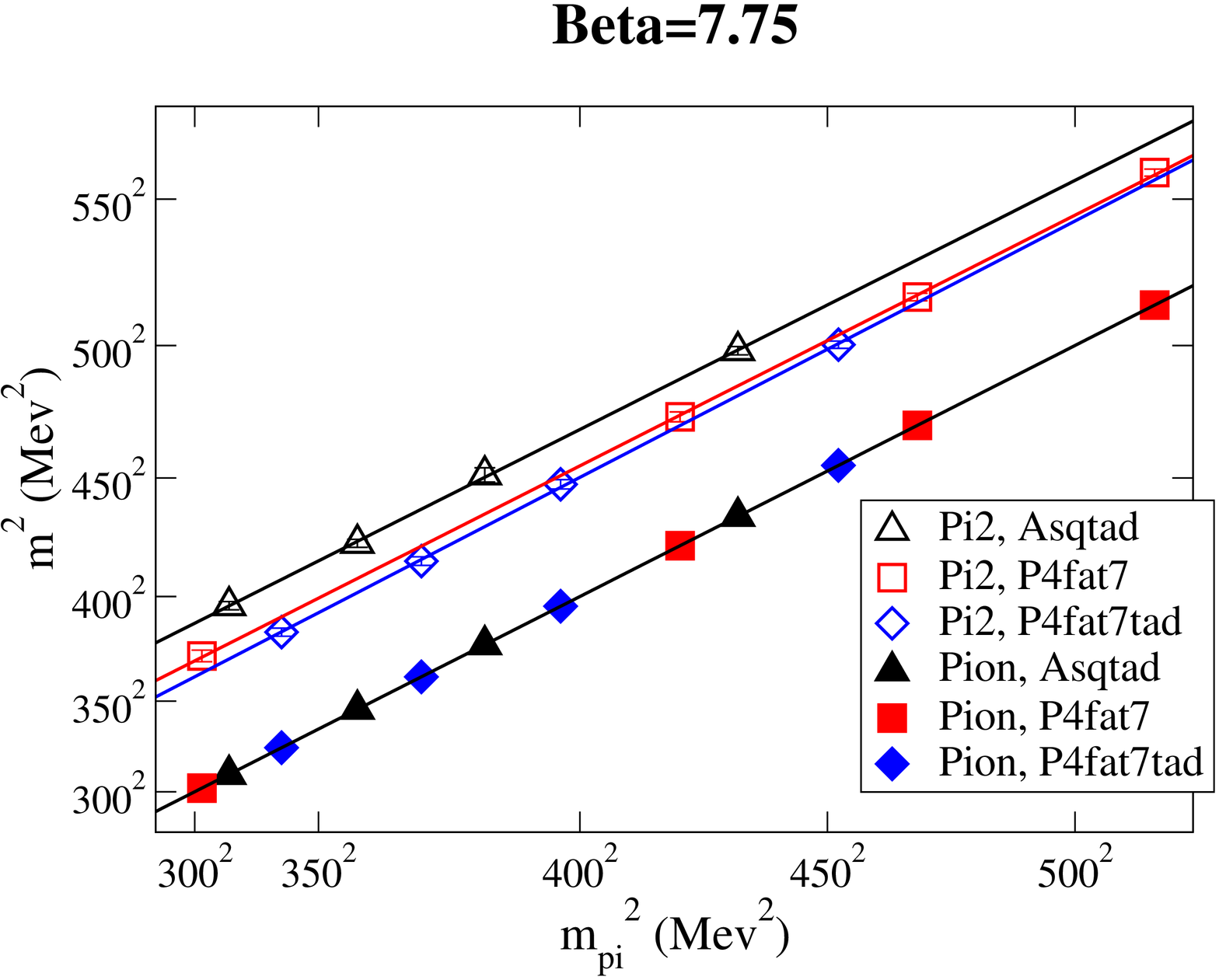}
  \includegraphics[angle=-90, width=.53\textwidth]{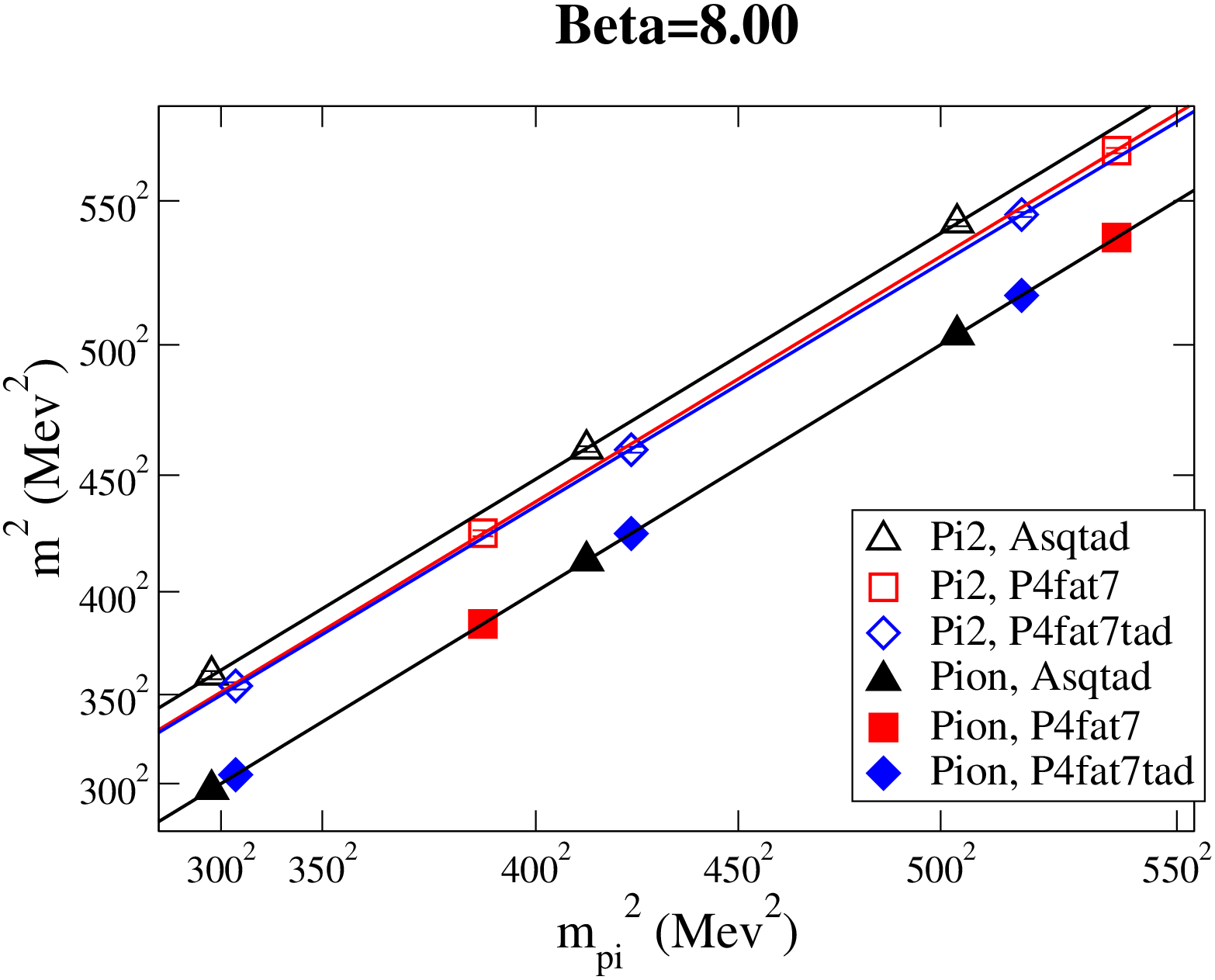}
  \includegraphics[angle=-90, width=.53\textwidth]{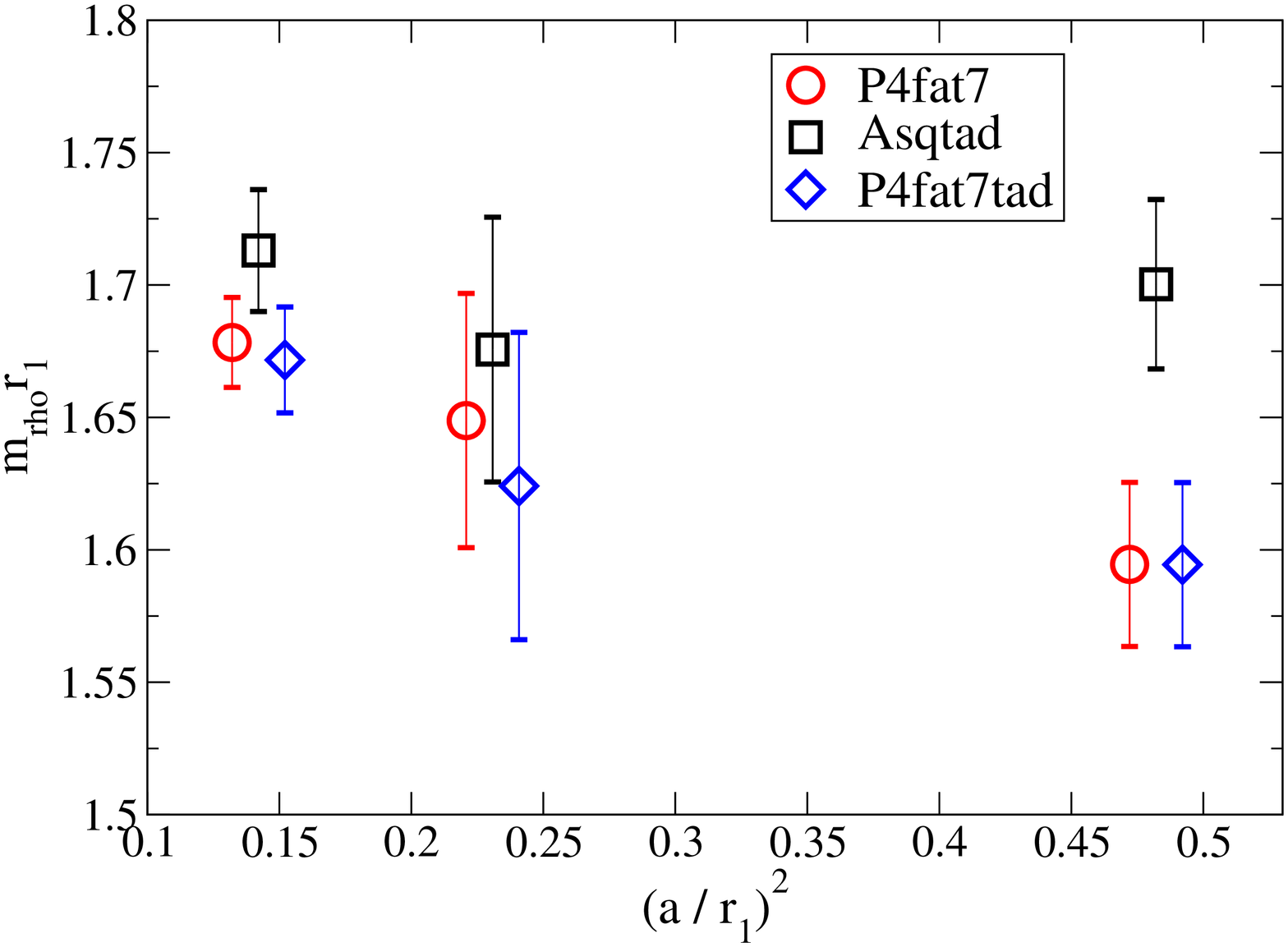}
  \caption{\label{plots}The figures on the top left, top right, and bottom 
    left show $m_{\pi_2}^2$ vs. $m_{\pi}^2$ for $\beta=7.40$, $\beta=7.75$,
    and $\beta=8.00$ respectively.  In our notation, $\pi_2$ denotes the scalar
    meson channel (non-Goldstone pion) and $\pi$ is the normal pseudoscalar
    channel.  The bottom right figure shows $m_{\rho}r_1$ vs. $a^2/r_1^2$ for 
    the various actions (data points offset for clarity).}
  \vspace{-4mm}
\end{figure}
Figure \ref{plots} show $m_{\pi}^2$ and $m_{\pi_2}^2$ plotted against 
$m_{\pi}^2$.  The straight line through the filled shapes represent 
$m_{\pi}^2$ for each of the three actions.  The other three lines through 
the unfilled shapes are chiral extrapolations for $m_{\pi_2}^2$.

As expected, the splitting between the two masses is largest on
the coarsest lattice, but diminishes as we go to smaller lattice spacing.
Also, we can see that the pion splitting is, in all cases, slightly
better for both P4fat7(red squares) and P4fat7tad(blue diamonds) 
as compared to Asqtad(black triangles).  Employing
tadpole improvement does seem to improve the pion mass splitting at the
coarser lattice spacings, but makes no difference for the finest lattice
point ($\beta=8.00$).

The bottom-right plot in Figure \ref{plots} also shows the dependence
of $m_{\rho}$ on lattice spacing.  For this plot, all masses are extrapolated
to the common physical point $m_{\pi}r_1 = .778$.  Then, $m_{\rho}$
is plotted against $a^2$, with all scales set in units of the string
tension scale $r_1$.  The black squares reproduce MILC results for the
Asqtad reasonably well.  The red octagons and blue diamonds show the scaling
dependence for the P4 actions.  P4fat7 and P4fat7tad give very similar answers,
showing an approximately 5\% change in $m_{\rho}$ over a factor of 2 in
lattice spacing.
\vspace{-2mm}

%%%%%%%%%%%%%%%%%%%%%%%%%%%%%%%%%%%%%%%%%%%%%%%%%%%%%%%%%%%%%%%%%%%%%%%%%

\section{Conclusions and Outlook}
In terms of the mass splitting between the non-Goldstone and the Goldstone
pion ($m_{\pi_2}-m_{\pi}$), both variants of the P4 action show
less flavor-symmetry breaking than the Asqtad action.  Tadpole improvement
also slightly improves this flavor splitting on the coarsest lattice,
but has almost no effect on the two finer lattices that were studied.  In
terms of the scaling of $m_{\rho}$, Asqtad does better, showing
little change from $a \approx .14$fm. to $a \approx .31$fm.  The P4 action
shows an approximately 5\% decrease over this same range.

To further study the scaling behavior of the P4 action, we plan to compare
these results with measurements using the unsmeared P4 action, to insure
that we see the expected improvement that comes from fat-link smearing.  
Also, it would be useful to look
at the entire set of pions whose mass is split by $\mathit{O}(a^2)$ effects.
This would provide a more complete picture of the flavor-symmetry violations
for each action.  The RBC-Bielefeld collaboration is now
using the P4 action in studying thermodynamic quantities.  Therefore, it
would also be valuable to study the scaling effects for correspondingly
coarser, dynamical lattices.
\vspace{-2mm}

\section*{Acknowledgements}
Thanks to Norman Christ, Frithjof Karsch, Peter Petreczky, 
Konstantin Petrov, and Christian Schmidt for useful and enlightening 
discussions.  Special thanks to Chulwoo Jung for managing the CPS++ 
software as well as assistance in implementing and testing the simulation 
code. 

We also thank Peter Boyle, Dong Chen, Norman Christ, Mike Clark, Saul Cohen, 
Calin Cristian, Zhihua Dong, Alan Gara, Andrew Jackson, Balint Joo, 
Chulwoo Jung, Richard Kenway, Changhoan Kim, Ludmila Levkova, Huey-Wen Lin, 
Xiaodong Liao, Guofeng Liu, Robert Mawhinney, Shigemi Ohta, Tilo Wettig, 
and Azusa Yamaguchi for the development of the QCDOC machine and its software. 
This development and the resulting computer equipment were funded
by the U.S. DOE grant DE-FG02-92ER40699, PPARC JIF grant PPA/J/S/1998/00756
and by RIKEN. This work was supported by DOE grant DE-FG02-92ER40699 and
we thank RIKEN, BNL and the U.S. DOE for providing the facilities essential
for the completion of this work.
\vspace{-2mm}

\bibliography{lat2005}

\providecommand{\href}[2]{#2}\begingroup\raggedright\begin{thebibliography}{1}

\bibitem{Orginos:1999cr}
{\bf MILC} Collaboration, K.~Orginos, D.~Toussaint, and R.~L. Sugar, {\it
  Variants of fattening and flavor symmetry restoration},  {\em Phys. Rev.}
  {\bf D60} (1999) 054503, [\href{http://xxx.lanl.gov/abs/hep-lat/9903032}{{\tt
  hep-lat/9903032}}].

\bibitem{Bernard:1999xx}
{\bf MILC} Collaboration, C.~W. Bernard {\em et~al.}, {\it Scaling tests of the
  improved kogut-susskind quark action},  {\em Phys. Rev.} {\bf D61} (2000)
  111502, [\href{http://xxx.lanl.gov/abs/hep-lat/9912018}{{\tt
  hep-lat/9912018}}].

\bibitem{Heller:1999xz}
U.~M. Heller, F.~Karsch, and B.~Sturm, {\it Improved staggered fermion actions
  for qcd thermodynamics},  {\em Phys. Rev.} {\bf D60} (1999) 114502,
  [\href{http://xxx.lanl.gov/abs/hep-lat/9901010}{{\tt hep-lat/9901010}}].

\bibitem{Lee:1999zx}
W.-J. Lee and S.~R. Sharpe, {\it Partial flavor symmetry restoration for chiral
  staggered fermions},  {\em Phys. Rev.} {\bf D60} (1999) 114503,
  [\href{http://xxx.lanl.gov/abs/hep-lat/9905023}{{\tt hep-lat/9905023}}].

\bibitem{Lepage:1992xa}
G.~P. Lepage and P.~B. Mackenzie, {\it On the viability of lattice perturbation
  theory},  {\em Phys. Rev.} {\bf D48} (1993) 2250--2264,
  [\href{http://xxx.lanl.gov/abs/hep-lat/9209022}{{\tt hep-lat/9209022}}].

\bibitem{Naik:1986bn}
S.~Naik, {\it On-shell improved lattice action for qcd with susskind fermions
  and asymptotic freedom scale},  {\em Nucl. Phys.} {\bf B316} (1989) 238.

\bibitem{Lepage:1998vj}
G.~P. Lepage, {\it Flavor-symmetry restoration and symanzik improvement for
  staggered quarks},  {\em Phys. Rev.} {\bf D59} (1999) 074502,
  [\href{http://xxx.lanl.gov/abs/hep-lat/9809157}{{\tt hep-lat/9809157}}].

\bibitem{Lepage:1997id}
P.~Lepage, {\it Perturbative improvement for lattice qcd: An update},  {\em
  Nucl. Phys. Proc. Suppl.} {\bf 60A} (1998) 267--278,
  [\href{http://xxx.lanl.gov/abs/hep-lat/9707026}{{\tt hep-lat/9707026}}].

\end{thebibliography}\endgroup

\end{document}